\documentclass[12pt]{article}

\usepackage{times}
\usepackage{graphicx}
\usepackage{amssymb}
\usepackage{amsthm}
\usepackage{amsmath}
\usepackage{dsfont}
\usepackage{bm}
\usepackage{mathrsfs}
\usepackage{bbold}
\usepackage{color}

\begin{document}
\title{\bf Questioning the recent observation of quantum Hawking radiation}
\author{Ulf Leonhardt\\
Weizmann Institute of Science,
Rehovot 761001,
Israel
}
\date{\today}

\maketitle

\begin{abstract}
A recent article [J. Steinhauer, Nat. Phys. {\bf 12}, 959 (2016)] has reported the observation of quantum Hawking radiation and its entanglement in an analogue black hole. This paper analyses the published evidence, its consistency with theoretical bounds and the statistical significance of the results. The analysis raises severe doubts on the observation of Hawking radiation.\\[3mm]
{\bf Key words} \\[3mm]
Analogues of gravity \\ Waves in moving fluids \\ Bose-Einstein condensates \\ Hawking radiation
\end{abstract}

\newpage

\section{Introduction}

Observations of Hawking radiation \cite{Hawking} in the laboratory \cite{Unruh} seem to have been vexed with problems. The very first of such observations \cite{Faccio} --- with intense light pulses in optical media, turned out to be inconsistent with theory \cite{Comment,Reply}. The observation of stimulated Hawking radiation of water waves \cite{Weinfurtner} was replicated  \cite{Germain0} and found to be anomalous scattering with and without a horizon \cite{Germain0}. The demonstration of black--hole lasing in Bose--Einstein condensates \cite{Stein} has been regarded as a fluid--mechanical instability different from lasing\cite{Caruso,Jacobson1,Jacobson2}, although the author { disagrees} \cite{SteinReply}. Nevertheless, these attempts of observing Hawking radiation in the laboratory have been tremendously fruitful, because their scientific debate has significantly advanced the subject. One would have learned much less from a perfect experiment right from the beginning. 

This paper analyzes one of the latest experimental demonstrations of Hawking radiation, an article \cite{Steinhauer} that states `Hawking radiation stimulated by quantum vacuum fluctuations has been observed in a quantum simulator of a black hole.' The paper scrutinises the published evidence and the methods of obtaining it. In doing so, it does not ask whether Hawking radiation {\it can} be observed \cite{Michel}, but rather whether it {\it has} been observed. The author of the article \cite{Steinhauer} had posted an informal response \cite{SteinhauerReply} to the criticism made in a previous version of this paper; here we also address the relevant points of this response \cite{ReplyNote}.

The article \cite{Steinhauer} in question reports on an experiment with Bose-Einstein condensates that are brought into motion by optical forces. An analogue of the event horizon is formed when the flow of the condensate $u$ exceeds the speed of sound $c$. Hawking's theory \cite{Hawking} applied to this case \cite{Unruh} predicts that the horizon creates sound quanta, phonons, from the fluctuations of the quantum vacuum. On each side of the horizon, the phonon population $\overline{n}$ is expected \cite{Hawking} to be equal and Planck-distributed 
%%%
\begin{equation}
\overline{n}=\frac{1}{\mathrm{e}^{\frac{\hbar\omega}{KT}}-1}
\label{planck}
\end{equation}
%%%
where $\omega=2\pi f$ denotes the circular frequency, $\hbar$ Planck's constant divided by $2\pi$ and $K$ Boltzmann's constant. The temperature $T$ is given \cite{Visser} by the relative velocity gradient at the horizon:
%%%
\begin{equation}
KT = \frac{\hbar}{2\pi} \left| \frac{\mathrm{d}(u-c)}{\mathrm{d}x}\right|_{\text{horizon}} \,.
\label{temp}
\end{equation}
%%%
Furthermore, the quantum particles are predicted \cite{Hawking} to be produced in maximally entangled pairs \cite{Barnett,Adamek} --- one phonon outside the horizon, the other inside. As one of the Hawking partners is beyond the horizon, one could never hope to observe this entanglement on astrophysical black holes, but in laboratory analogues one could. 

Hawking's Planckian prediction (\ref{planck}) is only valid in a regime of weak dispersion, as will be explained in Sec.~2. The experiment \cite{Steinhauer}, however, operates at the borderline between weak and strong dispersion, and the observed population distribution is clearly influenced by dispersion. Beyond a critical frequency, no radiation is measured. Yet the article \cite{Steinhauer} states that `thermal behaviour is seen at very low and very high energies.' Furthermore, the article \cite{Steinhauer} reports on correlations of Hawking partners beyond the critical frequency where there are no particles, which is impossible, as Sec.~3 is going to prove, unless the data are inconclusive, see Sec.~4. The data analysis of the article \cite{Steinhauer} is analized in Sec.~5 and conclusions are drawn in Sec.~6.

\section{Dispersion}
In laboratory analogues $c$ is dispersive: it depends on frequency or wavenumber $k=\omega/c$. A horizon is formed where the velocity $u$ reaches the group velocity $\partial ck/\partial k$. Due to dispersion the horizon is no longer sharply defined: it depends on wavenumber as well. Hawking's Planckian prediction (\ref{planck}) is only valid in a regime of weak dispersion \cite{Unschutz}. Sound in Bose-Einstein condensates at rest obeys the Bogoliubov dispersion relation \cite{SP}:
%%%
\begin{equation}
c^2= c_0^2 \left(1+\frac{\xi^2k^2}{2}\right)
\label{bogoliubov}
\end{equation}
%%%
where $c_0$ denotes the speed of sound for $k\rightarrow 0$ and the length $\xi=\hbar/(mc_0)$ quantifies the dispersion ($\xi$ is also the healing length of the condensate \cite{SP}). In condensates moving with velocity profiles $u$, sound waves experience the Bogoliubov dispersion in locally co--moving frames for the Doppler--shifted frequency:
%%%
\begin{equation}
\omega - uk = ck \,.
\label{dispersion}
\end{equation}
%%%
Figure \ref{stein3} shows measurements \cite{Steinhauer} of the wavenumbers in different regions of the condensate (taken pixel--by--pixel from the electronic version of Fig.~3 of the article \cite{Steinhauer}). One sees that, inside the horizon, there is a maximal frequency $\omega_c$ for waves propagating against the current. This is a simple consequence of Bogoliubov's relation (\ref{bogoliubov}): as the group velocity $\partial ck/\partial k$ of sound in the condensate increases with increasing $k$, there is a critical $k_\mathrm{peak}$ where the group velocity reaches the maximal velocity of the condensate in the experiment. Beyond the corresponding critical frequency $\omega_c$ there is no horizon. For $\omega \ll \omega_\mathrm{c}$ one still gets a Planck spectrum of Hawking particles if the dispersion is weak, {\it i.e.}\ if $\xi$ is significantly smaller than the characteristic length scale of the transition region where $u$ turns from subsonic to supersonic \cite{Unschutz,Macher}. As the Hawking temperature (\ref{temp}) is proportional to the relative velocity gradient at the horizon, most experimental analogues are forced to operate at the borderline between weak and strong dispersion, for maximising the particle yield. This is not necessarily a problem, but rather an opportunity to explore physics beyond Hawking's prediction (\ref{planck}). 

Figure \ref{stein5} shows the particle population versus wavelength inferred from measurements \cite{SteinTheory} (taken pixel--by--pixel from the electronic version of Fig.~5b of the article \cite{Steinhauer}). One clearly sees that beyond a certain wavenumber $k_\mathrm{c}$ the population vanishes within the error bars. According to the dispersion measurements shown in Fig.~\ref{stein3}, the value of $k_\mathrm{c}$ agrees with the wavenumber outside the horizon that corresponds to the critical frequency $\omega_\mathrm{c}$. These findings are consistent with a Hawking spectrum that deviates from the perfect Planck curve (\ref{planck}).

The article states \cite{Steinhauer}: `the Hawking distribution at low energies is thermal in the sense that the population goes like $1/\omega$'. This refers to the low-frequency limit of the Planck curve, the Rayleigh-Jeans limit:
%%%
\begin{equation}
\overline{n} \sim \frac{KT}{\hbar\omega}  \,.
\label{rayleigh}
\end{equation}
%%%
However, in the experiment \cite{Steinhauer} the length scale of the transsonic region is comparable with the scale $\xi$ of the dispersion. There is no guarantee that the constant $T$ in Eq.~(\ref{rayleigh}) has the meaning of a Hawking temperature (\ref{temp}) that is proportional to the velocity gradient. For example, an infinitely steep step from subsonic to supersonic speed would, according to Eq.~(\ref{temp}), create an infinite Hawking temperature, yet due to dispersion the population is finite and behaves for small $\omega$ like Eq.~(\ref{rayleigh}) as well \cite{Carusotto}. 

In the article \cite{Steinhauer}, a second method is used to calculate the Hawking temperature of the particle spectrum that does not suffer from this problem: the temperature is inferred from fitting the particle population obtained outside of the horizon with a Planck curve that is linearly brought to zero at $k_c$ (Fig.~\ref{stein5}). Here the deviation from the perfect Planck spectrum is taken into account with an {\it ad hoc} model, the linear descent to zero. The standard deviation of the fit (the square root of the sum of the squared differences between data and fit), 0.025, lies within the error bar, 0.028, of the data, but a simple linear fit would give a standard deviation of 0.039, which is only marginally worse than the fit used to infer the Hawking temperature. Furthermore, the accuracy of the fit is dominated by a single data point --- the one at  the left margin of the population data at $k_\mathrm{out}\xi_\mathrm{out} = 0.97$, because this point lies nearly exactly on the theory curve. Without this point, the Planckian {\it ad hoc} model had a standard deviation of $0.024$, which is worse than the standard deviation $0.015$ of the linear fit in this case.

No measurement of the velocity gradient at the horizon was reported \cite{Steinhauer}, so a quantitative comparison of the inferred temperature with the theoretical Hawking temperature of Eq.~(\ref{temp}) is not possible. Hence one cannot claim with certainty that the inferred $T$ is indeed a Hawking temperature.

\section{Entanglement}

Hawking radiation is predicted \cite{Hawking} to be entangled; each Hawking phonon that escapes from the horizon leaves a partner particle behind that drifts away on the other side. Moreover, Hawking radiation should be maximally entangled \cite{Barnett,Adamek} in a sense made precise below. The article \cite{Steinhauer} reports to have verified this prediction.

Quantum entanglement exists when the correlation strength exceeds a classical bound. In the article \cite{Steinhauer} the degree of entanglement is quantified by the non--separability condition \cite{Busch} 
%%%
\begin{equation}
\big|\langle \hat{b}_\mathrm{H}\hat{b}_\mathrm{P}\rangle\big|^2 > \overline{n}_\mathrm{H} \overline{n}_\mathrm{P} 
\label{criterion}
\end{equation}
%%%
where the $\hat{b}$ are annihilation operators; the index H refers to the Hawking particles outside the horizon and the index P to their partner particles inside. In classical statistical physics, the Cauchy--Schwarz inequality for probabilities implies that $\big|\langle b_\mathrm{H}b_\mathrm{P}\rangle\big|^2 \le \overline{n}_\mathrm{H} \overline{n}_\mathrm{P}$ \cite{Busch}, so violations of this bound are evidence for quantum entanglement. Note that the criterion of Eq.~(\ref{criterion}) is equivalent to the Peres-Horodecki criterion of entanglement for stationary and homogeneous modes \cite{SteinTheory}.

Maximally entangled states are the pure states of two modes with maximal entropy of the reduced density matrix, given a fixed energy \cite{Barnett}. For two modes of waves, such as the elementary excitations of Bose--Einstein condensates \cite{SP} of the experiment \cite{Steinhauer}, these are \cite{Barnett} the two--mode squeezed states \cite{LeoBook}. In this case, the populations of the Hawking partners are the same,
%%%
\begin{equation}
\overline{n}_\mathrm{H} = \overline{n}_\mathrm{P} = \overline{n} \,,
\label{pairs}
\end{equation}
%%%
and $\overline{n}$ { is} given by the Planck spectrum (\ref{planck}). { Furthermore, the joint quantum state of the Hawking mode and its partner is a two--mode squeezed state \cite{Barnett}} that obeys { \cite{Macher}}
%%%
\begin{equation}
\big|\langle \hat{b}_\mathrm{H}\hat{b}_\mathrm{P}\rangle\big|^2 = \overline{n}(\overline{n}+1) \,.
\label{heisenberg}
\end{equation}
%%%

Figure \ref{stein6} (taken pixel--by--pixel from the electronic version of Fig.~6a of the article \cite{Steinhauer} but displayed without the $S_0^2$ prefactor \cite{S0}) shows the correlations $\big|\langle \hat{b}_\mathrm{H}\hat{b}_\mathrm{P}\rangle\big|^2$ between the claimed Hawking partners inferred from measurements of the density--density correlations of the condensate \cite{SteinTheory} and $\overline{n}_\mathrm{H} \overline{n}_\mathrm{P}$ assuming $\overline{n}_\mathrm{H}= \overline{n}_\mathrm{P}$, as Eq.~(\ref{pairs}) predicts. The figure shows no entanglement { according to criterion (\ref{criterion})} for $k_\mathrm{in}\xi_\mathrm{in}<1.4$ inside the horizon that corresponds (Fig.~\ref{stein3}) to $k_\mathrm{out}\xi_\mathrm{out}<1.1$ outside the horizon where the agreement with the Planck curve (\ref{planck}) is best (Fig.~\ref{stein5}). It does display entanglement for medium $k$, but then it goes on to show correlations beyond the critical $k_\mathrm{peak}$ where no Hawking particles were observed, which, if one takes the population curve (Fig.~\ref{stein5}) literally, is impossible. One easily derives \cite{Adamek} from the Cauchy-Schwarz inequality for the correlation $\langle \hat{b}_\mathrm{H}\hat{b}_\mathrm{P}\rangle$ of Hawking and partner particles
%%%
\begin{equation}
\big|\langle \hat{b}_\mathrm{H}\hat{b}_\mathrm{P}\rangle\big|^2 \le \overline{n}_\mathrm{H} (\overline{n}_\mathrm{P} +1) 
\,,\quad
\big|\langle \hat{b}_\mathrm{H}\hat{b}_\mathrm{P}\rangle\big|^2 \le \overline{n}_\mathrm{P} (\overline{n}_\mathrm{H} +1) \,.
\label{bounds}
\end{equation}
%%%
For maximally entangled two--mode states the bound of Eq.~(\ref{bounds}) is saturated, as Eqs.~(\ref{pairs}) and (\ref{heisenberg}) show. Relation (\ref{heisenberg}) is called \cite{Steinhauer} the Heisenberg limit (although it does not originate \cite{Adamek} from Heisenberg's uncertainty relation). The bounds (\ref{bounds}) are valid for all quantum states, regardless of the specific values of the populations $\overline{n}_\mathrm{H}$ and $\overline{n}_\mathrm{P}$. In particular, the correlation must vanish for vanishing $\overline{n}_\mathrm{H}$ or $\overline{n}_\mathrm{P}$: no correlation exists without population, yet the article \cite{Steinhauer} shows correlations there (Fig.~\ref{stein6}).

The article \cite{Steinhauer} states one must convolve the population `with the $k$--distribution of the outgoing modes near $k_\mathrm{peak}$'. The rationale for the convolution is the following: as the experiment lasts for a finite time $\tau$ the frequencies $\omega$ have the uncertainty $\Delta\omega\sim1/\tau$. As one cannot discriminate between the frequencies within $\Delta\omega$, even in--principle, the population must be averaged over the wavenumber range $\Delta k$ that corresponds to $\Delta\omega$ via the dispersion relation. The averaging is done by the convolution. The dispersion curve $\omega(k)$ for phonons inside the horizon reaches a maximum at $k_\mathrm{peak}$, and so $\mathrm{d}k/\mathrm{d}\omega$ tends to $\infty$ there; small uncertaintes in frequencies result in large wavenumber uncertainties, which increases the error bars of $k$ near $k_\mathrm{peak}$ (Fig.~\ref{stein3}). There are two contributions to the error bars: the finite extensions of the observation regions and the finite observation time; for  $k\sim k_\mathrm{peak}$ the latter dominates. 

Here we model the point--spread function of the measurement by a Gaussian \cite{Cutoff} (Fig.~\ref{convolution}). We convolve with a Gaussian the theoretical population curve $\overline{n}$ in the Heisenberg limit $\overline{n}(\overline{n}+1)$. The theoretical curve is the Planckian fit linearly brought to zero of the population data (insert of Fig. \ref{stein5}). The resulting convolution curve (Fig.~\ref{convolution}) fits the last four data points of the correlation data (Fig.~\ref{stein6}) if the standard deviation $\sigma$ of the Gaussian is set to the constant $1.21$ for $k$ in units of $1/\xi$. This $\sigma$ lies slightly above the maximal error bar $0.94$ obtained from the dispersion measurements (Fig.~\ref{stein3}). Our curve (Fig.~\ref{convolution}) is virtually indistinguishable from the corresponding curve of the article \cite{Steinhauer} obtained with the actual point--spread function, which shows that the Gaussian is an excellent model \cite{ConvolutionNote}. 

The agreement of the correlation data with the Heisenberg limit for large wavenumbers --- beyond $k_\mathrm{peak}$ --- is seen \cite{Steinhauer} as the strongest evidence for the entanglement of Hawking radiation. However, the { convolution} with maximal { uncertainty} is only justified in the vicinity of $k_\mathrm{peak}$; { only there the small frequency uncertainty $\Delta\omega$ due to the finite time of the experiment results in a large wavenumber uncertainty $\Delta k$. The} dispersion measurements show (Fig.~\ref{stein3}) { that only near $k_\mathrm{peak}$ the wavenumber uncertainty rises.} Taking the variation of $\sigma$ into account \cite{Remark} moves the curve of $\overline{n}(\overline{n}+1)$ below the correlation curve (Fig.~\ref{convolution}): the observed correlations violate the { fundamental bounds of Eq.~(\ref{bounds})} for large wavenumbers. 

For smaller wavenumbers, the correlations drop significantly below the prediction of maximal entanglement, Eq.~(\ref{heisenberg}). The article \cite{Steinhauer} states that here `the Hawking pairs are less correlated than expected or they are produced in smaller quantities'. The latter, however, would contradict the population measurements. The two indications for Hawking radiation, the Planck spectrum for small wavenumbers and the { reported} maximal entanglement for large wave numbers, are not consistent. 

\section{Uncertainties}

\begin{figure}
\begin{center}
\includegraphics[width=22pc]{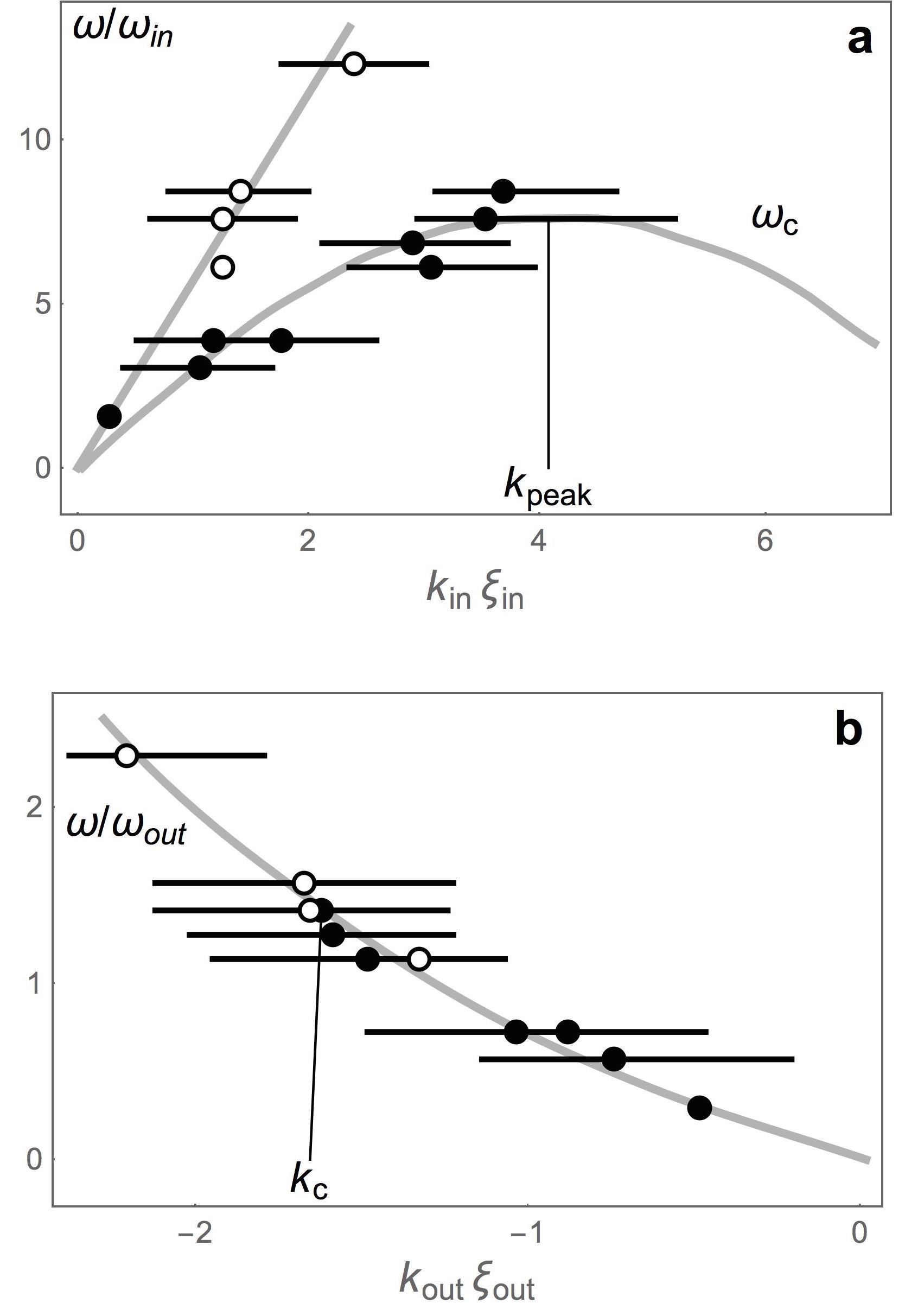}
\caption{
\small{
Dispersion measurements. {\bf a} inside the horizon: wavenumbers $k_\mathrm{in}$ corresponding to frequencies in units with $\hbar\omega_\mathrm{in} = mc_\mathrm{in}^2$ where $m$ is the atomic mass; $\xi_\mathrm{in}$ is the dispersion/healing length of Eq.~(\ref{bogoliubov}). The gray curves show the solutions of the dispersion relation, Eq.~(\ref{dispersion}), with fitted parameters. One sees two branches, one (full dots) of Hawking waves trying to escape but not succeeding, the other (open dots when distinguishable) of waves propagating with the flow. At $k_\mathrm{peak}$ the Hawking waves are fast enough to reach the flow velocity and are no longer trapped;  there is no horizon beyond the critical frequency $\omega_c$. {\bf b} outside the horizon: wavenumbers \cite{Scaling} versus frequencies (notation analogous to {\bf a}). Here $k_\mathrm{c}$ is the wavenumber that corresponds to $\omega_c$. The data points and error bars were taken from Fig.~3 of the article \cite{Steinhauer}. { Note that the error bars correspond to the half width at half maximum of the observed resolution \cite{Steinhauer}; one has to divide them by $1.2$ for getting the standard deviation $\sigma$ \cite{SteinhauerReply}.}
}
\label{stein3}}
\end{center}
\end{figure}

\begin{figure}
\begin{center}
\includegraphics[width=22pc]{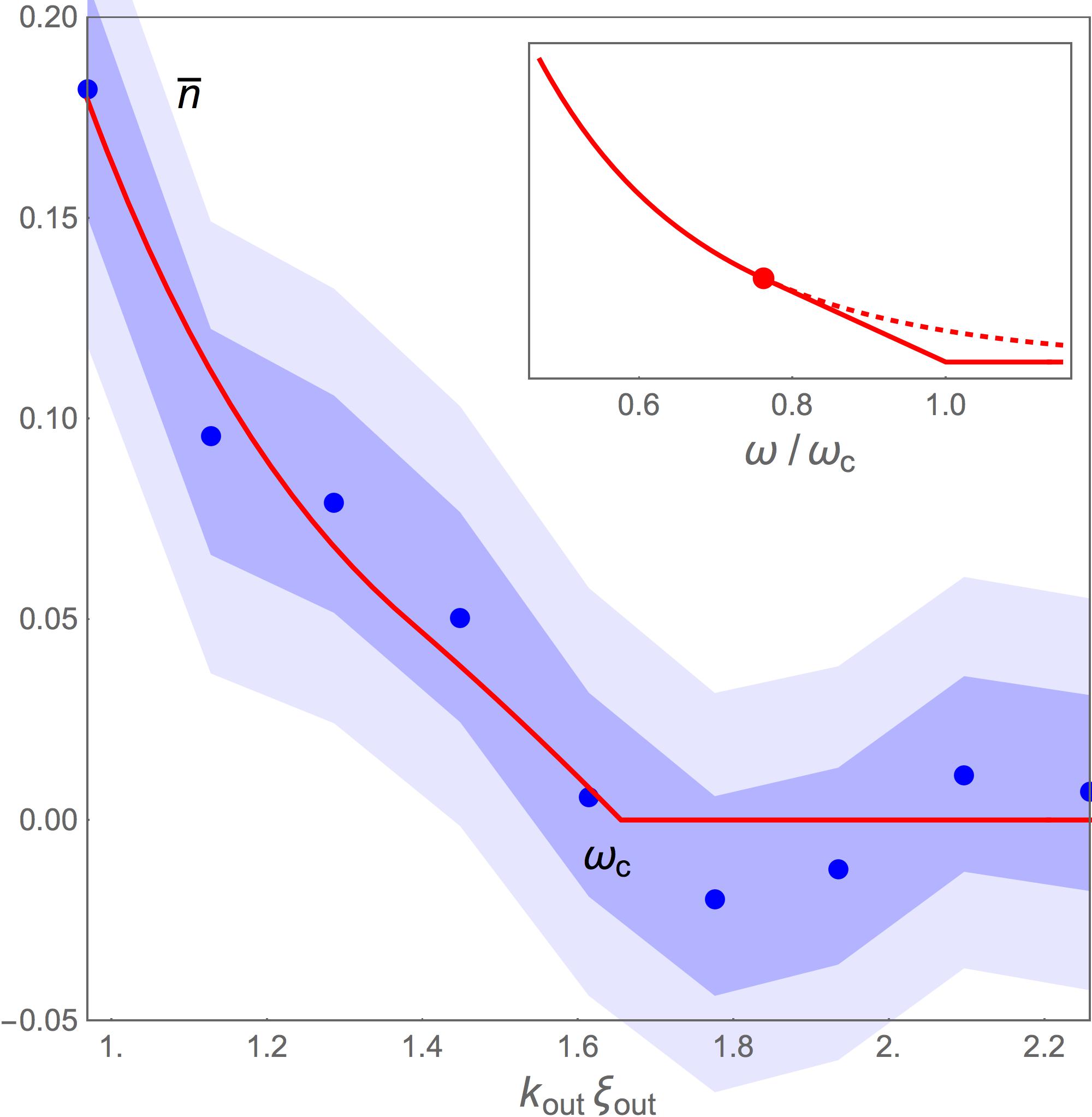}
\caption{
\small{
Population of outgoing particles. The blue dots represent the phonon populations $\overline{n}$ obtained from experiment \cite{Steinhauer} at wavenumbers $k_\mathrm{out}$ outside the horizon (notation as in Fig.~\ref{stein3}). The points are surrounded by their vertical error bars; the dark--blue area represents $1\sigma$ and the light--blue area $2\sigma$ (where $\sigma$ means the standard deviation). The data were taken from Fig.~5b of the article \cite{Steinhauer}. The population is zero within the error bars for the data points corresponding to frequencies beyond $\omega_\mathrm{c}$ (Fig.~\ref{stein3}). The red line shows a fit with a Planck curve linearly brought to zero at $\omega_\mathrm{c}$, as the insert illustrates. There the dotted line is the Planck curve continued beyond the red point of deviation. The data is consistent with the dispersion measurements (Fig.~\ref{stein3}) and theoretical expectations, but the spectrum is clearly not Planckian. 
}
\label{stein5}}
\end{center}
\end{figure}

\begin{figure}
\begin{center}
\includegraphics[width=22pc]{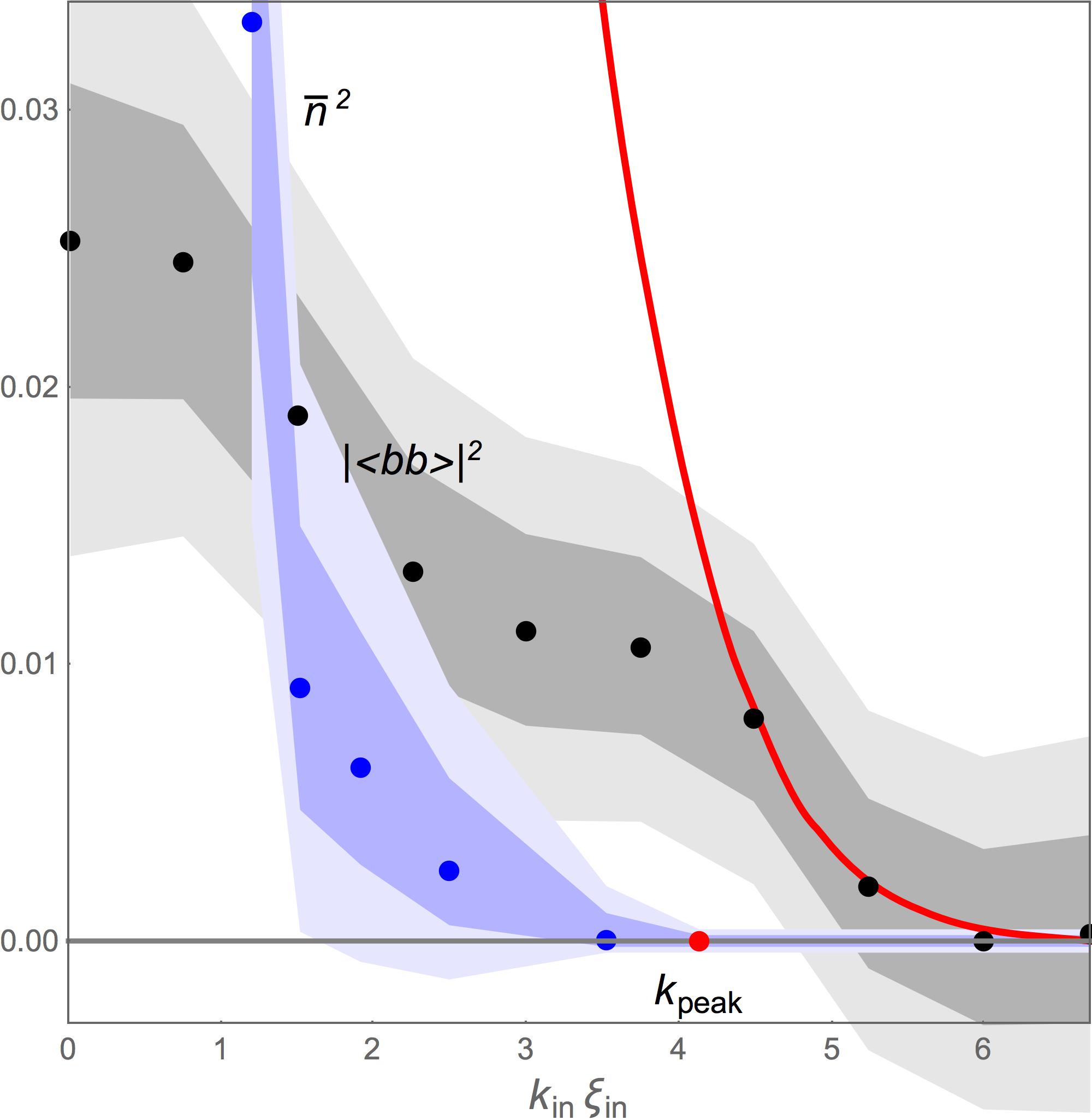}
\caption{
\small{
Entanglement. The Hawking phonons are entangled with their partners if the correlations $|\langle \hat{b}_\mathrm{H}\hat{b}_\mathrm{P}\rangle|^2$ (black dots) lie above the populations squared $\overline{n}^2$ (blue dots), assuming the populations of Hawking and partner particles are the same. As in Fig.~\ref{stein5} the points are surrounded by their vertical error bars; the darker areas indicate $1\sigma$, the lighter areas $2\sigma$. The data (from Fig.~6a of Ref.~\cite{Steinhauer} without $S_0^2$ prefactor \cite{S0}) are shown versus wavenumber $k_\mathrm{in}$ inside the horizon; the $k_\mathrm{out}$ of the populations (Fig.~\ref{stein5}) are transformed via $\omega$ into $k_\mathrm{in}$ (Fig.~\ref{stein3}). The populations vanish beyond $k_\mathrm{peak}$ (red dot) but not the correlations. The red curve shows the Heisenberg limit $\overline{n}(\overline{n}+1)$ with $\overline{n}$ obtained by convoluting the population curve with a Gaussian (Fig.~\ref{convolution}).
}
\label{stein6}}
\end{center}
\end{figure}

\begin{figure}
\begin{center}
\includegraphics[width=22pc]{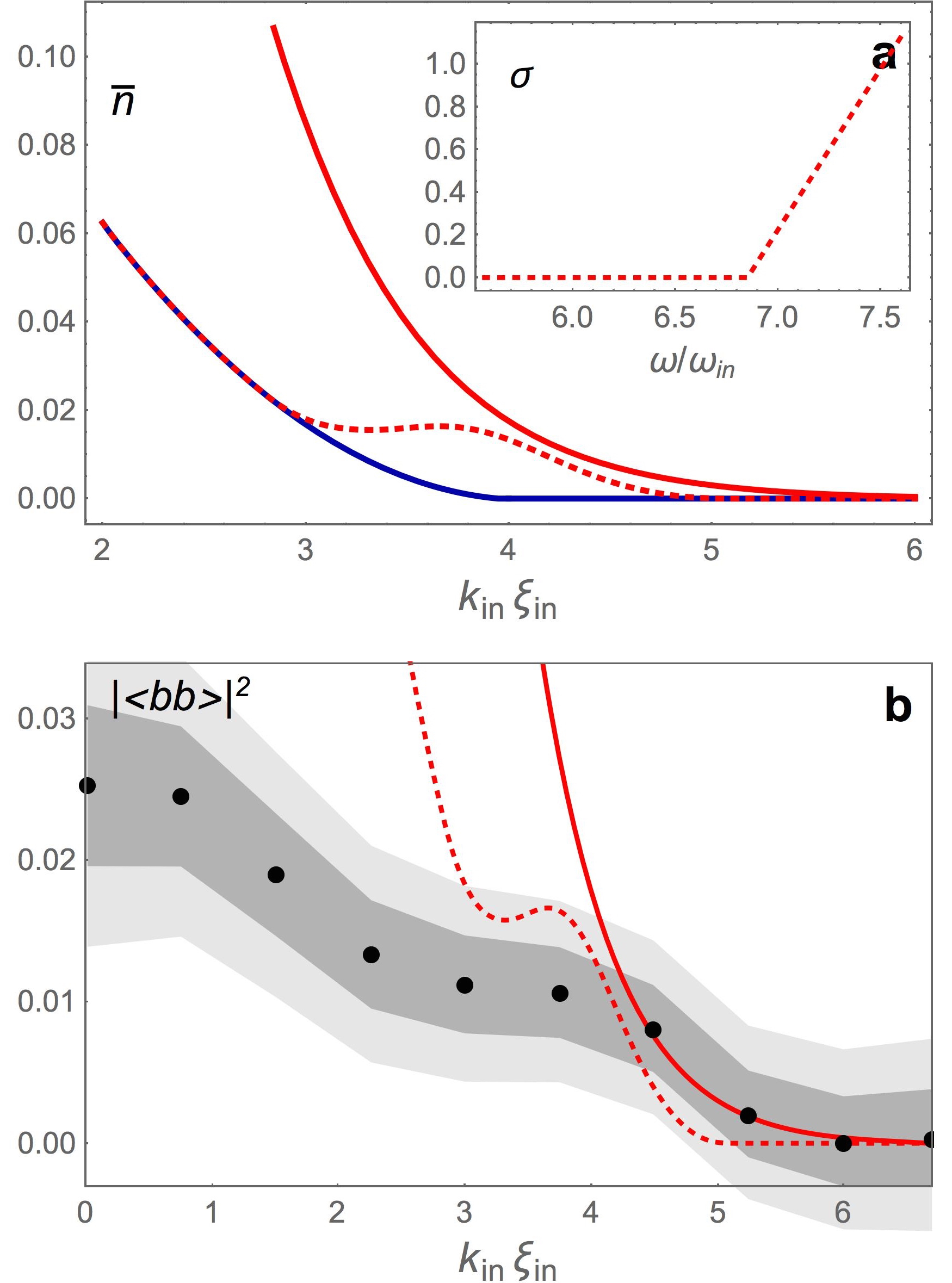}
\caption{
\small{
Convolution. {\bf a}: 
The fitted population curve (dark blue, from Fig.~\ref{stein5}) is represented as function of $k_\mathrm{in}\xi_\mathrm{in}$ and convoluted with a Gaussian to produce the red curve for $\overline{n}$ that gives the Heisenberg limit $\overline{n}(\overline{n}+1)$ of Fig.~\ref{stein6} in agreement with the article \cite{Steinhauer}. The standard deviation of the Gaussian was set to the constant $1.21$. The dotted curve shows the population curve convoluted with variable standard deviation $\sigma$ displayed in the insert that reflects the actual uncertainty in $k_\mathrm{in}\xi_\mathrm{in}$ taken from the dispersion measurements \cite{Remark} (Fig.~\ref{stein3}). {\bf b}: comparison of the Heisenberg limits $\overline{n}(\overline{n}+1)$ with fixed (red) and variable (red, dotted) convolution with the particle correlations \cite{Steinhauer} (black dots with uncertainty regions from Fig.~\ref{stein6}). Beyond the critical wavenumber $k_\mathrm{peak}$ the correlation curve tends to lie above the corrected Heisenberg limit (dotted line), which violates the fundamental bounds of Eq.~(\ref{bounds}). 
}
\label{convolution}}
\end{center}
\end{figure}

\begin{figure}
\begin{center}
\includegraphics[width=22pc]{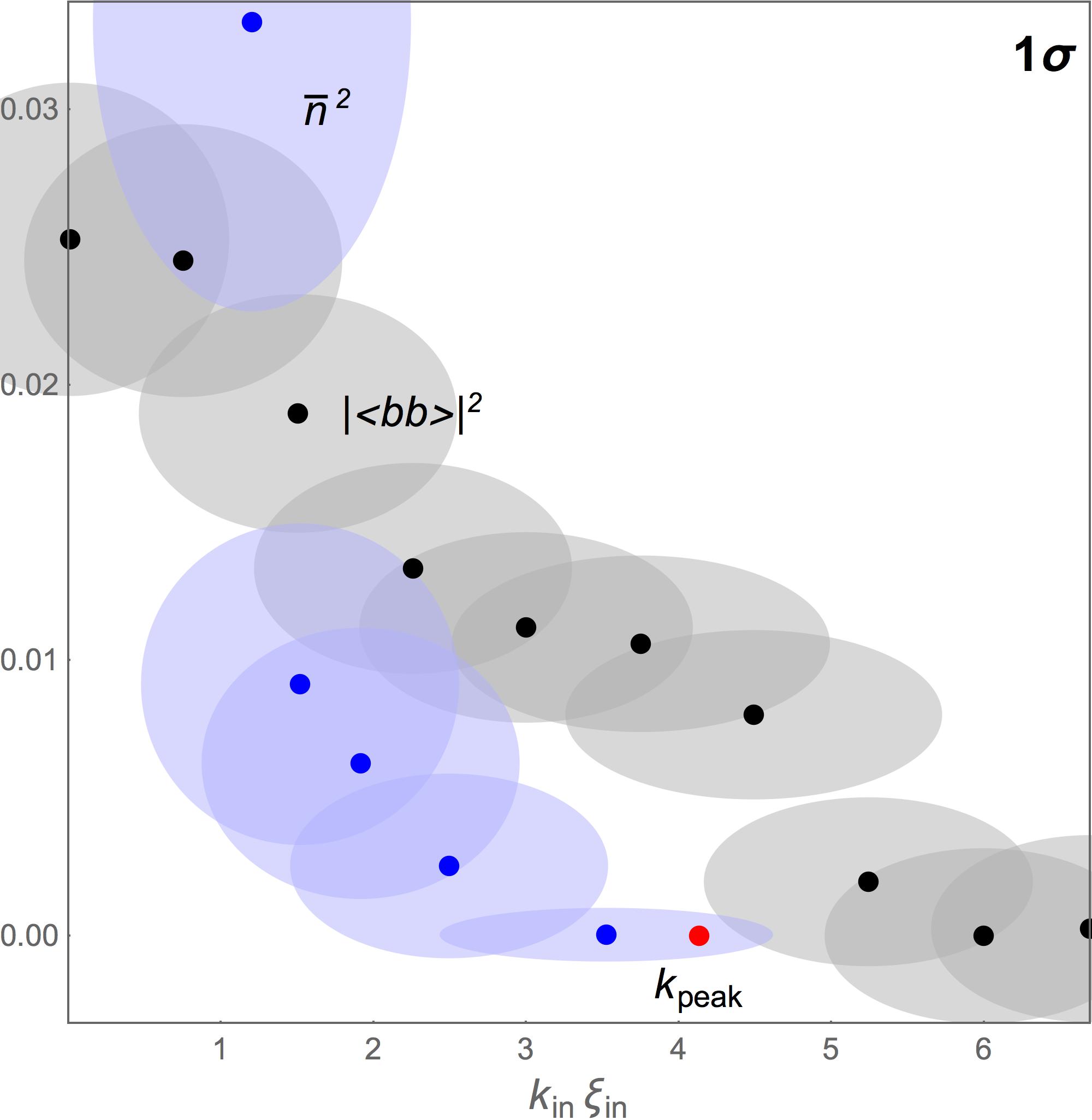}
\includegraphics[width=22pc]{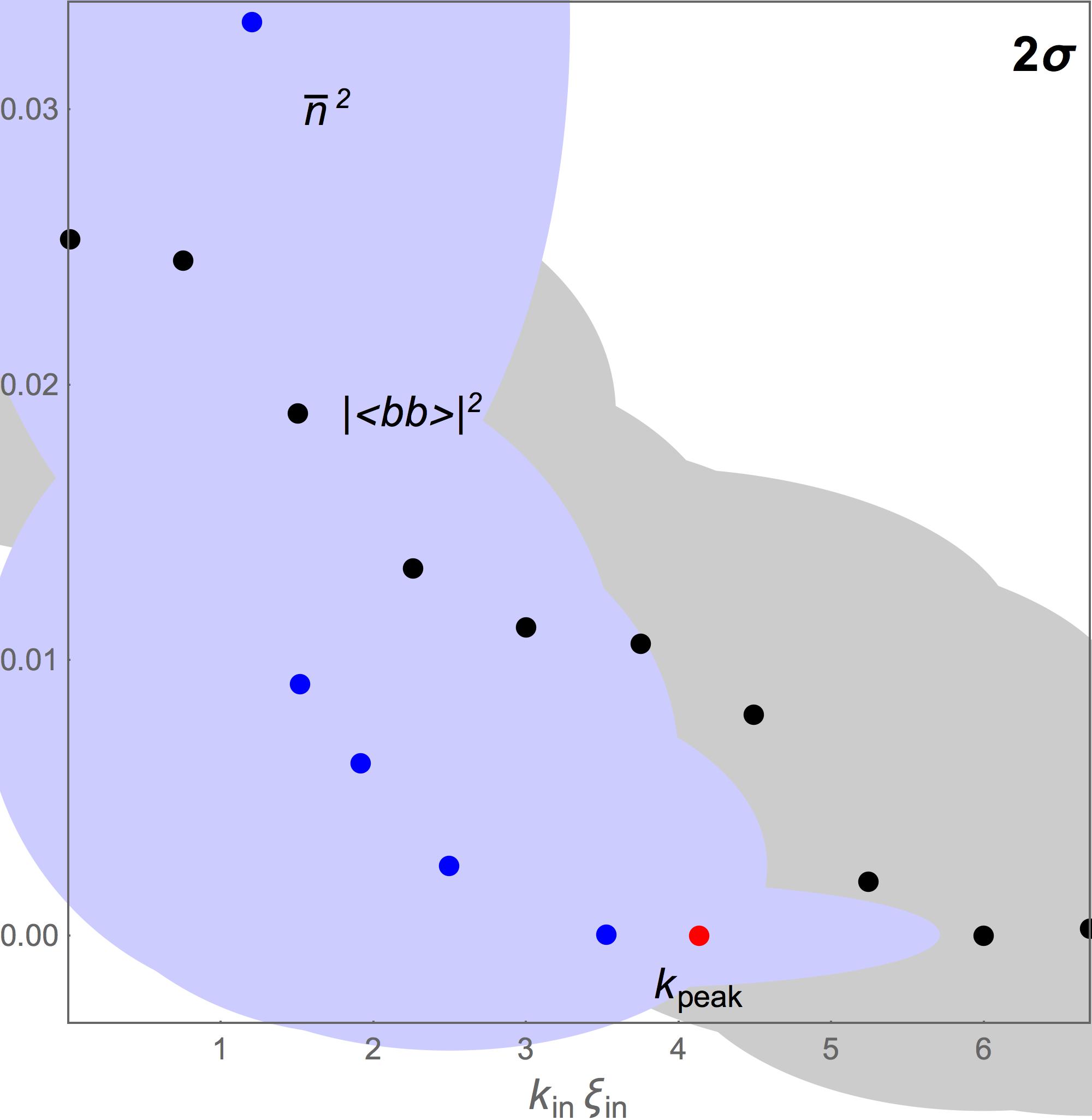}
\caption{
\small{
Entanglement with full error bars. The data of Fig.~\ref{stein6} is shown with full error regions (ellipses). The uncertainties in the variables $k_\mathrm{in}\xi_\mathrm{in}$ are obtained from Eq.~(\ref{sigma}) and the dispersion measurements (Fig.~\ref{stein3}) \cite{Scaling}. $\bm{1\sigma}$: the correlations $|\langle \hat{b}_\mathrm{H}\hat{b}_\mathrm{P}\rangle|^2$ (black dots, gray ellipses) are distinguishable from the populations squared $\overline{n}^2$ (blue dots, lightblue ellipses). $\bm{2\sigma}$: the curves are indistinguishable. 
}
\label{errors}}
\end{center}
\end{figure}

Perhaps these inconsistencies disappear if all uncertainties in the experiment are taken into account (Fig.~\ref{errors}). So far, it was assumed that the populations and correlations are functions of perfectly sharp wavenumber, except in the Gaussian smoothing of the population curve for fitting the Heisenberg limit $\overline{n}(\overline{n}+1)$ to the tail of the correlation data. However, the { measured} wavenumbers do carry uncertainties (Fig.~\ref{stein3}), primarily due to the finite sizes of the two observation regions (one outside, one inside the horizon). They are the result of Fourier analysis \cite{Steinhauer,SteinTheory} and so their accuracy is limited in finite regions. Note that we do not regard the wavenumber uncertainties as due to statistical errors; they are assumed to be solely given by the resolution of the optical measurement and the effect of the finite time of the experiment, in contrast to what Ref.~\cite{SteinhauerReply} states. Note also that  the correlations $|\langle \hat{b}_\mathrm{H}\hat{b}_\mathrm{P}\rangle|^2$ and $\overline{n}_\mathrm{H}\overline{n}_\mathrm{P}$ depend on two wavenumbers, $k_\mathrm{H}$ and $k_\mathrm{P}$ ($k_\mathrm{out}$ and $k_\mathrm{in}$ in Fig.~\ref{stein3}). This remains true if $\overline{n}_\mathrm{H}=\overline{n}_\mathrm{P}$ is assumed; the product $\overline{n}_\mathrm{H}(k_\mathrm{H})\,\overline{n}_\mathrm{P}(k_\mathrm{P})=\overline{n}_\mathrm{H}(k_\mathrm{H})\,\overline{n}_\mathrm{H}(k_\mathrm{H}(k_\mathrm{P}))$ is still a function of the two variables $k_\mathrm{H}$ and $k_\mathrm{P}$ with uncertainties. The correlation data is plotted as function of one variable, $k_\mathrm{P}=k_\mathrm{in}$, where the other variable $k_\mathrm{H}$ is related to $k_\mathrm{P}$ via the dispersion curves (Fig.~\ref{stein3}) \cite{Scaling}. For taking both uncertainties into account, the uncertainty $\sigma$ of the wavenumber  $k_\mathrm{in}$ used for plotting is { given by} the convolution of the two individual { point--spread functions that describe the measurement resolution. Assuming them to be Gaussian \cite{ConvolutionNote}, the total variance $\sigma^2$ is the sum of the two individual variances,
%%%
\begin{equation}
\sigma^2=\sigma_\mathrm{H}^2+\sigma_\mathrm{P}^2 \,.
\label{sigma}
\end{equation}
%%%
One obtains the uncertainties from the dispersion measurements (Fig.~\ref{stein3}) by dividing the error bars --- the half width at half maximum \cite{Steinhauer} --- by a factor of $1.2$ \cite{SteinhauerReply}, and interpolates them \cite{Scaling}. Figure \ref{errors} shows the result: if the uncertainties of the wavenumbers are taken into account the data of the correlations is only distinguishable from the data of the populations squared for one standard deviation, the two data sets melt into each other for $2\sigma$ \cite{Twosigma}. So, either one accepts inconsistencies in the data or the data become insignificant.

\section{Analysis}

An important part of the critical analysis of experimental data is the analysis of the methods used to obtain them, for finding out whether there is bias in the methods. Experimental high--energy physics has the highest standards in identifying and eliminating bias, because there experiments cannot be easily repeated at different facilities. Experimental observations of Hawking radiation in Bose--Einstein condensates can be repeated in principle, but in practice it is difficult --- and unwise --- to repeat exactly the same experiment of the article \cite{Steinhauer}; it is much more productive to use more sophisticated methods, both in the preparation and in the measurement technique. But that leaves the question open whether the pioneering experiment \cite{Steinhauer} has achieved its objective. It is clear that quantum Hawking radiation {\it can} be observed \cite{Michel}, but {\it has} it been observed?

Let us therefore analyse the assumptions made in the data analysis of the article \cite{Steinhauer}. For quantifying the degree of entanglement, the correlation was compared with the population of only one of the Hawking partners (the one outside the horizon) and not also with the population of the other. The article \cite{Steinhauer} contains a hint \cite{Hint} that also the population of the Hawking partners was measured, but the data were not published. Reference~\cite{SteinhauerReply} is silent about this point. For analysing the entanglement, it was assumed that $\overline{n}_\mathrm{H}$ and $\overline{n}_\mathrm{P}$ were the same. However, this was part of Hawking's prediction, Eq.~(\ref{pairs}), and cannot be taken for granted. For example, this assumption is not guaranteed if the initial state of the excitations differs from the vacuum state \cite{Calculation}, which is not impossible, given that the analogue of the event horizon was made by a non--stationary process in the experiment \cite{Steinhauer} --- moving an optical barrier over the condensate. The theory \cite{Michel} does not take this into account.

Consider now the method \cite{SteinTheory} of obtaining the particle correlations from the experimental data. This method is based on the Fourier analysis of the density--density correlations obtained from averaging the measured density profiles along the positions $x_\mathrm{H}$ and $x_\mathrm{P}$ over the $4,600$ experimental runs \cite{Steinhauer}. Yet the full Fourier transform of the density--density data was not used in the article \cite{Steinhauer}. The particle correlations were inferred from subsets of the data integrated along lines parallel to a line of expected correlations (see Fig.~4 of the article \cite{Steinhauer}), a line found by optimisation. This method selects the Fourier--components of the density--density correlations evaluated within a subset of the data that would be consistent with the expected particle correlations \cite{SteinTheory}. All other Fourier--components were ignored in the published article \cite{Steinhauer}. Although this would give the Hawking correlations if they are there, it does not allow a comparison with the level of the other Fourier components: it does not discriminate between signal and noise, nor between signal and background. Reference~\cite{SteinhauerReply} is also silent about this point, although a Fourier analysis of the full density--density data is easily performed and should be reported.

The subsets of the density--density data were chosen, because they display a typical feature expected from Hawking radiation in Bose--Einstein condensate \cite{Moustache}: a thin anti--diagonal band--shaped line highlighted in green in the electronic version of Fig.~4a of the article \cite{Steinhauer}. This characteristic line corresponds to a drop in the density--density correlation (Fig.~4b of the article \cite{Steinhauer}). The anti--correlation in density is associated with the correlation of particles emitted from the horizon \cite{Moustache}, one moving away from it, the other being dragged with the flow. These particles could be the ones of Hawking radiation from the quantum vacuum, but the characteristic pattern of the anti--diagonal line is also drawn by anything else originating from the horizon and moving with the group velocity of waves in the condensate. It depends on the quantitative details of the density--density drop in comparison with the background whether it is indeed Hawking radiation originating from the quantum vacuum. The article \cite{Steinhauer} argues that this is the case. Yet in calculating only the Fourier components orthogonal to the characteristic line and ignoring all other Fourier components one biases the data analysis and does not get the signal to noise.

The presence of statistical noise might explain why the experimental results appear to agree well with theory at the margins of $k_\mathrm{in}$ (Figs.~\ref{stein5} and \ref{stein6}). The particle population lies remarkably well on the Planck curve of Eq.~(\ref{planck}) at the lowest $k_\mathrm{in}$ value, whereas the tail of the particle correlations agrees almost perfectly with the Heisenberg limit of Eq.~(\ref{heisenberg}) for large $k_\mathrm{in}$, provided the population $\overline{n}$ is given by the convolution with the full point--spread function. As we have discussed in Sec.~3, the latter was an incorrect theory, because only the effect due to the finite observation time should be taken into account here. However, noisy data may produce such features, in particular at the margins.

Note that} the method of extracting correlations \cite{SteinTheory} --- even from the complete density--density data --- involves another assumption: it assumes that only the Hawking {pairs} are generated in accelerating the condensate beyond the speed of sound, but no other excitations in the fluid. As justification, the article states \cite{Steinhauer}: `the neglected terms represent correlations between widely separated phonons on opposite sides of the horizon with different frequencies.' However, while this is true for the discrimination between in--modes (Fig.~\ref{stein3}a) and out--modes (Fig.~\ref{stein3}b), co-- and counter--propagating waves may mix: Figure~\ref{stein3}a shows the wavenumbers and frequencies for the two possible modes of excitations inside the horizon (represented by filled and open circles in the figure): the Hawking--partner waves attempting to travel against the flow and waves traveling with the flow. One sees that their wavenumbers and frequencies are similar, so they could have contaminated the data. In any case, neglecting the influence of extra excitations on the density--density correlation amounts to an assumption that has not been independently verified in the article \cite{Steinhauer}. 

\section{Conclusions}

This paper has analysed the evidence for the observation of quantum Hawking radiation and its entanglement in the recent article \cite{Steinhauer} and found several problems. {\it First}, the observed spectrum of Hawking radiation is not { fully} Planckian, although the article \cite{Steinhauer} {states} it to be. {\it Second}, particle correlations are observed without particles being present, unless, {\it third},  wavenumber uncertainties reduce the { confidence in the data} to one standard deviation, {\it i.e.}\ to insignificance. {\it Fourth}, no measurement of the population of Hawking partners was reported. {\it Fifth}, the method of inferring particle correlations from the measured density--density correlations {was biased towards} the expectation { of quantum} Hawking radiation { and the signal--to--noise was not reported.} 

While the experimental data of the article \cite{Steinhauer} seem valid, although incomplete, the conclusions { appear questionable}, especially the { statement \cite{Steinhauer}} of having observed entanglement with a confidence of $90/16=5.7\sigma$ \cite{SteinhauerReply}, which reduces to the order of $1\sigma$ { after an analysis of the uncertainties involved}. The 750 GeV bump in the LHC data had a confidence of $2.1\sigma$, but still turned out to be a coincidence \cite{LHC}. One cannot claim that quantum Hawking radiation and its entanglement has been observed by the standards of a discovery.  What has been observed is a matter of debate.

To close this paper on a positive note, I would like to advocate an unbiased view on the physics of the event horizon in laboratory analogues, and related phenomena, because only in this case real discoveries can be made. We should be open to surprises. One of the recent surprises in this field was the experimental observation \cite{Germain} that ripples in flowing water behave very similar to quantum Hawking radiation when the water exceeds the speed of the waves at a horizon. Hidden in the ripples of the water surface, the analogue of Hawking radiation was revealed {by space--time Fourier analysis} and correlations between Hawking waves were seen. There the slight turbulence of the water played the role of the quantum vacuum, stimulating Hawking radiation from classical fluctuations. The fact \cite{Germain} that this is possible was indeed surprising. This paper \cite{Germain} also sets a high bar in the quality of experimental data and the standards of data analysis and theoretical understanding. Another recent highlight --- also with water --- is the observation of superradiance in a vortex flow \cite{Torres}. While classical fluids like water cannot show quantum entanglement, it is still remarkable that something as mundane and simple as flowing water can reveal many of the features of something as mysterious and subtle as Hawking radiation \cite{Hawking}.

\section*{Acknowledgements}

I thank
Yael Avni,
Jonathan Drori,
Itay Griniasty
and Yuval Rosenberg for discussions,
and Jeff Steinhauer for answering our many questions on his article.
This work was supported by the European Research Council and the Israeli Science Foundation.

%%%
%%%

%%%
%%%

%%%
%%%

%%%
%%%

%%%
%%%


\begin{thebibliography}{99}
\bibitem{Scaling}
The scales $\xi_\mathrm{in}$ and $\xi_\mathrm{out}$ of the wavenumbers in the figures are related to each other by $\xi_\mathrm{in}/\xi_\mathrm{out}=c_\mathrm{out}/c_\mathrm{in}$, because $\xi=\hbar/(mc_0)$ \cite{SP}. In the experiment \cite{Steinhauer} $c_\mathrm{out}=0.57\mathrm{mm}/\mathrm{s}$ and $c_\mathrm{in}=0.25\mathrm{mm}/\mathrm{s}$.
\bibitem{S0}
In the article \cite{Steinhauer} the particle correlations (Fig.~6a of Ref.~\cite{Steinhauer}) are displayed with prefactors $S_0^2$ that depend \cite{Steinhauer} on the Bogoliubov--de Gennes wavefunctions of the elementary excitations \cite{SP}. However, by comparing $S_0^2 \overline{n}^2$ (also in Fig.~6a of Ref.~\cite{Steinhauer}) with $\overline{n}^2$ (Fig.~5b of Ref.~\cite{Steinhauer}) we realized that $S_0^2$ is nearly constant. Therefore, for keeping the paper self--contained without introducing Bogoliubov--de Gennes wavefunctions, we assumed  $S_0^2$ to be constant and displayed both $|\langle \hat{b}_\mathrm{H}\hat{b}_\mathrm{P}\rangle|^2$ and $\overline{n}^2$ without $S_0^2$.
\bibitem{Cutoff}
The convolution of the Planck curve (\ref{planck}) as function of $\omega(k_\mathrm{in})$ with a Gaussian would diverge due to the pole (\ref{rayleigh}) unless it is cut off; here we cut it off for $k_\mathrm{in}\xi_\mathrm{in}<0.05$.
\bibitem{ConvolutionNote} 
The good agreement of the red curve in Fig.~\ref{convolution}b with the solid-grey curve in Fig.~6a of Ref.~\cite{Steinhauer}, obtained by convolution with the actual point--spread function of the optical apparatus, shows that the Gaussian is a good model for describing the measurement resolution, in contrast to what is stated in Ref.~\cite{SteinhauerReply}.
\bibitem{Hint}
The section on measuring the population of Hawking radiation in the article \cite{Steinhauer} mentions a correction factor of 2.2 for the population outside the horizon and a factor of 2.3 for the population inside, which indicates that the population $\overline{n}_\mathrm{P}$ was measured and not only $\overline{n}_\mathrm{H}$.
\bibitem{Calculation}
The process of Hawking radiation \cite{LeoBook} transforms the initial annihilation operators $\hat{a}_1$ and $\hat{a}_2$ into 
$\hat{b}_\mathrm{H}=\hat{a}_1\cosh\zeta+\hat{a}_2^\dagger\sinh\zeta$ and 
$\hat{b}_\mathrm{P}=\hat{a}_2\cosh\zeta+\hat{a}_1^\dagger\sinh\zeta$
with $\tanh^2\zeta=\exp(-\hbar\omega/KT)$. 
Suppose the initial state is $|\alpha\rangle_1|0\rangle_2$ with $|\alpha\rangle$ being a coherent state \cite{LeoBook}. One finds that the final state is always entangled. However, in the article \cite{Steinhauer}, it would only appear to be entangled for $|\alpha|^2<\exp(-\hbar\omega/KT)$, as only then the correlation exceeds $\overline{n}_\mathrm{H}^2$. Note that $\overline{n}_\mathrm{H}\neq\overline{n}_\mathrm{P}$ here, which would still be consistent with the published data \cite{Steinhauer}.
\end{thebibliography}
\end{document}